\documentclass[11pt,a4paper]{emulateapj}

\usepackage{longtable}

\usepackage[usenames]{color}

\def\comment#1{}
\def\cblock#1{}

\usepackage[normalem]{ulem}

\def\crossout#1{}

\def\CIVdbl{{\rm C}\,{\sc iv}$\,\lambda\lambda 1548,1550$}
\def\MgIIdbl{{\rm Mg}\,{\sc ii}$\,\lambda\lambda 2796,2803$}
\def\NVdbl{{\rm N}\,{\sc v}$\,\lambda\lambda 1238,1242$}  
\def\OVIdbl{{\rm O}\,{\sc vi}$\,\lambda\lambda 1031,1037$}
\def\SiIVdbl{{\rm Si}\,{\sc iv}$\,\lambda\lambda1394,1403$}
\def\AlIIIdbl{{\rm Al}\,{\sc iii}$\,\lambda\lambda1855,1863$}

\def\FeIIlinB{{\rm Fe}\,{\sc ii}$\,\lambda 2383$}
\def\FeIIlinR{{\rm Fe}\,{\sc ii}$\,\lambda 2600$}
\def\CIIlin{{\rm C}\,{\sc ii}$\,\lambda\lambda 1335$}
\def\SiIIlin{{\rm Si}\,{\sc ii}$\,\lambda\lambda 1304$}
\def\SiIIdbl{{\rm Si}\,{\sc ii}$\,\lambda\lambda 1527,1533$}
\def\OIlin{{\rm O}\,{\sc i}$\,\lambda 1302$}

\def\AlIII{\hbox{{\rm Al}\,{\sc iii}}}

\def\CIII{\hbox{{\rm C}\,{\sc iii}}}
\def\CIV{\hbox{{\rm C}\,{\sc iv}}}
\def\CV{\hbox{{\rm C}\,{\sc v}}}

\def\HI{\hbox{{\rm H}\,{\sc i}}}

\def\Lya{\hbox{{\rm Ly}\,$\alpha$}}

\def\FeII{\hbox{{\rm Fe}\,{\sc ii}}}

\def\MgII{\hbox{{\rm Mg}\,{\sc ii}}}

\def\NV{\hbox{{\rm N}\,{\sc v}}}

\def\SiIII{\hbox{{\rm Si}\,{\sc iii}}}
\def\SiIV{\hbox{{\rm Si~}\,{\sc iv}}}

\def\kms{\hbox{km~s$^{-1}$}}

\def\a{$\alpha$}
\def\s{$\sigma$}
\def\lb{$\lambda$}

\def\lsim{\mathrel{\rlap{\lower4pt\hbox{\hskip1pt$\sim$}}
    \raise1pt\hbox{$<$}}}                
\def\gsim{\mathrel{\rlap{\lower4pt\hbox{\hskip1pt$\sim$}}
    \raise1pt\hbox{$>$}}}                

\begin{document}

\title{Discovery of the Transition of a Mini-Broad Absorption Line
  into Broad Absorption in the SDSS quasar
  J115122.14+020426.3}\label{sec:def}

\author{Paola~Rodr\'iguez~Hidalgo\altaffilmark{1,2}, Michael
  Eracleous\altaffilmark{1}, Jane Charlton\altaffilmark{1}, Fred
  Hamann\altaffilmark{3}, Michael Murphy\altaffilmark{4}, Daniel
  Nestor\altaffilmark{5}}

\altaffiltext{1}{Department of Astronomy and Astrophysics, The
  Pennsylvania State University, 525 Davey Lab, University Park, PA
  16802, USA}

\altaffiltext{2}{Department of Physics and Astronomy, 128 Petrie
  Science and Engineering Building, York University, 4700 Keele
  Street, Toronto, Ontario, M3J 1P3, Canada}

\altaffiltext{3}{Department of Astronomy, University of Florida,
  Gainesville, FL 32611, USA}

\altaffiltext{4}{Center for Astrophysics and Supercomputing, Swinburne
  University of Technology, PO Box 218, Hawthorn, Victoria 3122,
  Australia}

\altaffiltext{5}{Department of Physics \& Astronomy, 430 Portola
  Plaza, Box 951547, University of California Los Angeles, CA 90095,
  USA}

\begin{abstract}

\bigskip

We present the detection of a rare case of dramatic strengthening in
the UV absorption profiles in the spectrum of the quasar
J115122.14+020426.3 between observations $\sim$~2.86 years apart in
the quasar rest-frame. A 2001 spectrum from the Sloan Digital Sky
Survey (SDSS) shows a \CIV\ ``mini-broad'' absorption line
(FWHM=$1,220\;$\kms) with a maximum blueshift velocity
$\sim9,520\;$\kms, while a later spectrum from the Very Large
Telescope (VLT) shows significantly broader and stronger absorption
line, with a maximum blueshift velocity of $\sim12,240\;$\kms, that
qualifies as a broad absorption line. A similar variability pattern is
observed in two additional systems at lower blueshifted velocities and
in the Ly\a\ and \NV\ transitions as well. One of the absorption
systems appears to be resolved and shows evidence for partial covering
of the quasar continuum source ($C_f\sim$~0.65), indicating a
transverse absorber size of, at least, $\sim6\times 10^{16}\;$cm. In
contrast, a cluster of narrower {\CIV} lines appears to originate in
gas that fully covers the continuum and broad emission line
sources. There is no evidence for changes in the centroid velocity of the 
absorption troughs. This case suggests that, at
least some of the absorbers that produce ``mini-broad'' and broad
absorption lines in quasar spectra do not belong to intrinsically
separate classes. Here, the ``mini-broad'' absorption line is most
likely interpreted as an intermediate phase before the appearance of a
broad absorption line due to their similar velocity. While the current
observations do not provide enough constraints to discern among the
possible causes for this variability, future monitoring of multiple
transitions at high resolution will help achieve this goal.

\end{abstract}
\keywords{galaxies: active -- quasars:general -- quasars:absorption lines.}

\section{Introduction} 
\label{sec:1}

Outflows are fundamental constituents of Active Galactic Nuclei
(AGN). They are detected in a substantial fraction of AGNs through
their absorption-line signatures (e.g., broad, blueshifted resonance
lines in the UV and X-ray bands) and could be ubiquitous, if the
absorbing gas subtends a small solid angle to the central continuum
(and broad emission line) source (e.g. \citealt{Crenshaw99};
\citealt{Reichard03}; \citealt{Hamann04}; \citealt{Trump06};
\citealt{Nestor08}; \citealt{Dunn08}; \citealt{Ganguly08} and
references therein). Outflows have been invoked as a regulating mechanism in order to explain the
correlation between the black hole masses ($M_{\bullet}$) and the
masses of the stellar spheroids of their host galaxies
\citep[$M_{\rm bulge}$; e.g.,][]{Gebhardt00, Merritt01, Tremaine02}. The
evolutionary models developed to explain this relation invoke energy
and momentum ``feedback'' by outflows from the accreting supermassive
black hole onto the gas in the host galaxy \citep[e.g.,][]{Silk98,
  Springel05, DiMatteo05, Hopkins06}. The same AGN outflows have also
been suggested to distribute heavy elements into the intergalactic
medium \citep[e.g.,][]{Scannapieco04, Germain09}.

Blueshifted UV resonance absorption lines (e.g., \CIVdbl) are often
used as signposts of outflows. These lines are classified based on
their widths, as follows. Broad Absorption Lines (BALs), with widths
of several thousands of km~s$^{-1}$, and {\it intrinsic} Narrow
Absorption Lines (intrinsic NALs)\footnote{These lines have been
  termed ``intrinsic'' to indicate that they arise in gas that is
  related to the quasar central engine. They are distinct from narrow
  absorption lines that arise in intervening structures such as
  intervening galaxies or the intergalactic medium.}, with widths less
than a few hundred km~s$^{-1}$, are the most commonly studied classes
\citep[see, for example,][]{Weymann81, Turnshek84, Foltz86, Weymann91,
  Aldcroft94, Reichard03, Vestergaard03, Trump06}. Absorption lines
with intermediate widths, called ``mini-BALs'', have not been studied
as extensively \citep[e.g.,][] {Turnshek88, Januzzi96, Hamann97a,
  Telfer98, Churchill99, Ma02, Yuan02, Narayanan04, Misawa07b,
  Gibson09a}, and their nature remains poorly understood. 

Mini-BALs and BALs could, for example, trace absorbers with different
properties (i.e., different densities or ionic column densities that
may be attributed to the sizes of the absorbing parcels of gas or their
ionization states). Perhaps mini-BALs and BALs probe different regions of
the same gas flow where the physical conditions are different, which
would mean that both types of lines could be seen in the same type of quasar
but along different lines of sight. Yet another possibility is that
mini-BALs and BALs represent different stages in the time evolution of
a non-steady gas flow. The first of the above scenarios can be tested
through a better characterization of the physical properties of
mini-BALs and comparison to those of BALs.  The possibility that BALs
and mini-BALs can transform into each other can be tested through
monitoring observations. Variations of the absorption lines caused by
changes in the ionization of the gas yield constraints on the gas
density via the recombination time. Similarly, variability caused by
motion of parcels of gas across the line of sight provide information
about the density distribution, hence the ``granularity'' and phase
structure of the gas.  Thus, several monitoring campaigns have been
carried out aimed to characterize variability of BALs
(e.g., \citealt{Barlow93}; \citealt{Lundgren07}; \citealt{Gibson08};
\citealt{Capellupo11}). Serendipitous discoveries of variability have also contributed to this goal (e.g.,
\citealt{Ma02}; \citealt{Hamann08}; \citealt{Krongold10};
\citealt{Vivek12}).  These studies of variability can help us
understand better the geometry and location of the absorbing gas, as
well as its physical conditions (structure, sizes of gas parcels,
variability of the gaseous flow resulting from instabilities, etc.),
which could be used to test current theoretical models
\citep[i.e.,][]{Murray95, Proga00, Proga04, Proga12}.

Here we present a study of the dramatic variability of the
\CIV\ mini-BAL in the radio-quiet quasar J115122.14+020426.3 (hereafter J1151+0204). This quasar
was first observed in the Sloan Digital Sky Survey (SDSS) from which
its magnitude and redshift were reported to be $g=19.1$ and
$z_{\rm em}=2.401$ \citep{Schneider03}.  During our systematic study of
absorption features in the spectra of the $\sim$2200 brightest SDSS
quasars (Rodr\'iguez Hidalgo et al. in prep), we found a
\CIV\ mini-BAL in the spectrum of J1151+0204 at $z_{\rm abs}=2.296$ with FWHM of
$1,220\;${\kms} and a maximum blueshift velocity of
$\sim9,520\;$\kms. In an unrelated study of \MgII\ absorbers we came upon
an archival spectrum of this quasar taken with the Ultraviolet and
Visual Echelle Spectrograph at the Very Large Telescope (VLT/UVES). A
comparison of the two spectra showed the transformation of the
\CIV\ mini-BAL into a BAL with a maximum blueshift velocity of
$\sim12,240\;$\kms; this is a rare example of such a transformation. We
also detect variable absorption lines of other ionic species in the
same system, and the variability of other absorption systems
outflowing at lower velocity. In \S\ref{sec:2} of this paper we
present the data while in \S\ref{sec:3} we describe the analysis of
the spectra (including the normalization of the continuum and the
identification of the absorption lines) and present the
measurements. Finally, in \S\ref{sec:4} we consider further the
implications of our particular results and discuss what we can learn in general about the properties of
quasar outflows from variable absorption systems.

We adopt a redshift for J1151+0204 of $z_{\rm em} =2.399$, which
we determined by measuring the centroid of the peak of the
\CIII]$\lambda$1909 emission line in the UVES spectrum. Previous
  redshift determinations from the SDSS spectrum yielded $z_{\rm em}
  =2.401$ \citep{Schneider03} and $z_{\rm em}=2.409$
  \citep{Hewett10}. The former value is based on multiple emission
  lines, some of which may be contaminated by associated absorption
  lines, while the high-ionization lines may be blueshifted relative
  to the systemic redshift \citep{Shen07}. The latter value is based
  on cross-correlation of the {\CIII}] line profile with a template
    and may be affected by the {\AlIII}$\lambda$1857 and
    {\SiIII}]$\lambda$1892 lines. Relative velocities computed with
      our redshift and that of \citet{Hewett10} differ by
      $\sim900\;$\kms, but this does not alter any of our conclusions.
      Throughout this paper, the central wavelengths of absorption troughs, and thus absorption redshifts,
are determined through the apparent optical-depth-weighted mean of the profile,
as described in \citet{Churchill01}.

\section{Observations, Archival Spectra, and Basic Properties of J1151+0204}
\label{sec:2}

The spectrum J1151+0204 was first obtained as part of the Sloan Digital Sky Survey
in 2001 May 20 with a total exposure time of 3978~s and under ``excellent
conditions''.
It covers the range 3800--9200~{\AA} (in the observer's frame) at a
signal-to-noise ratio (S/N) of 16.4 pixel$^{-1}$ in the i band and 5.5
pixel$^{-1}$ at the shortest wavelengths. The resolving power is
$R\sim \lambda/\Delta \lambda \sim$2000 (corresponding to a velocity
resolution of $\sim$150 {\kms}; \citealt{Adelman-McCarthy08}).

The UVES spectrum was taken on three consecutive nights, on 2008
March 10--12, as part of a program on ultra-strong MgII absorbers
(VLT/UVES proposal 080.A-0795(A). Principal Investigators: Nestor,
Pettini, \&\, Rao). The total exposure time was 26,297~s, although two
different cross-disperser settings were employed (DICHR1 and DICHR2,
observations with exposure times of 12,615 and 13,682~s,
respectively). The S/N is not constant throughout the spectrum,
but it always exceeds 15 pixel$^{-1}$, which is adequate for our
purposes. The wavelength range covered is 3595--9466 {\AA}. The
resolving power is $R\sim45,000$ (corresponding to a velocity
resolution of 6.7~\kms).

\begin{figure}
\begin{center}
\includegraphics[width=9cm]{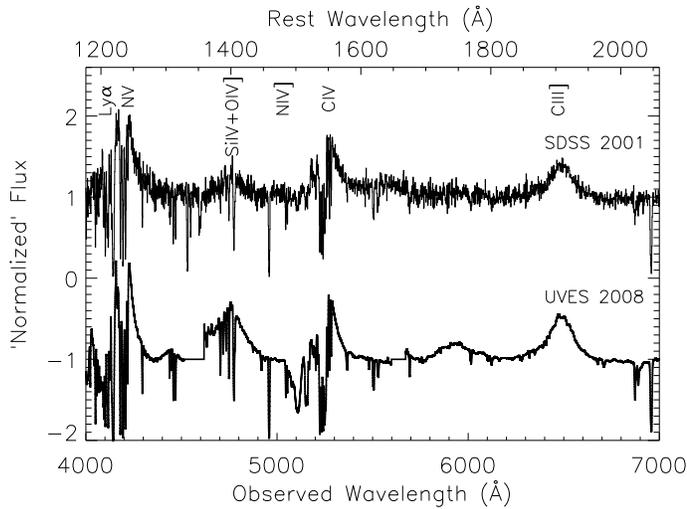}
{\vspace{0.5cm}} 
\caption[Figure1]{SDSS (upper trace) and UVES (lower trace) spectra of
  the quasar J1151+0204 taken in 2001 and 2008,
  respectively. These spectra illustrate the S/N attained and the
  emission and absorption features present in the spectra.  For the
  purposes of this illustration we used a polynomical or a power law
  to normalize the continuum in different segments of each
  spectrum. The UVES spectrum was then smoothed to the resolution of
  the SDSS spectrum for easier comparison and shifted downwards for
  clarity.  There are easily discernible gaps between orders in the
  UVES spectrum at wavelengths of 4500--4600~\AA\ and 5600--5700~\AA.}
\label{fig0}
\end{center}
\end{figure}

\begin{figure}
\begin{center}
\includegraphics[width=9cm]{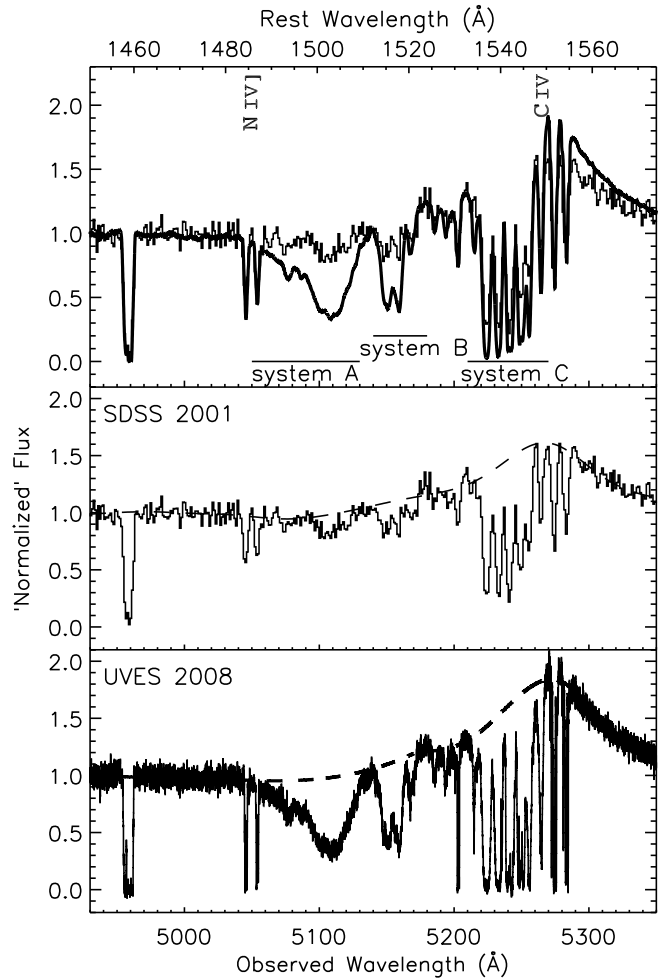}
{\vspace{0.5cm}} 
\caption[Figure1]{Enlarged view of the {\CIV} region of the spectra
  shown in Figure~\ref{fig0}. Top panel: The SDSS spectrum is represented by the
  thin/noisy line and the smoothed UVES spectrum by the thick line. The locations of
  common emission lines are labelled. The absorption
  systems discussed in the text are also marked. Middle panel and bottom panel show the SDSS spectrum and the UVES spectrum, respectively, with the effective continuum we used to normalize them. Notice how we followed a conservative approach in both cases but even more so in the case of the UVES spectrum where the {\CIV} emission line might be larger than what we suggest.}
\label{fig1}
\end{center}
\end{figure}

\comment{I moved the first 3 paragraphs of section 3 into this
  section. I also expanded the first paragraph to describe the new
  Figure~0.}

In Figure~\ref{fig0} we display the normalized SDSS and UVES
spectra. For the purposes of this illustration, we have normalized both
continua to unity (using a power-law fit to line-free regions of the
SDSS spectrum and a polynomial fit to those regions in the UVES
spectrum) and have smoothed the  UVES spectrum to the resolution of the
SDSS spectrum. The locations of common emission lines are
labelled and the rest-frame wavelengths here and elsewhere in the paper
are defined relative to the redshift $z_{em}=2.399$ (see \S\ref{sec:1}). 
This Figure shows the S/N attained in the spectra and the
main emission and absorption features present.  The appearance of the
\CIV\ BAL between 2001 and 2008 is obvious even in a cursory
inspection of the spectra. The appearance of a \Lya\ BAL can also be
discerned upon close inspection.

In Figure~\ref{fig1}, we show an expanded view of the \CIV\ region of
the SDSS and UVES spectra from Figure~\ref{fig0}, where we mark and
label the absorption-line systems discussed in this paper.  The
absorption trough in the 2001 SDSS spectrum at $\sim5108\;$\AA\ has
FWHM$=1,220\;$\kms\ and qualifies as a {\CIV} mini-BAL\footnote{In the survey of \CIV\ mini-BALs in SDSS spectra \citep[Rodr\'iguez Hidalgo et al.\ in prep]
{RodriguezHidalgo07, RodriguezHidalgo09} we found this absorption trough and categorized it as a candidate \CIV\ mini-BAL; no other plausible identification was found.} at
$z_{\rm abs}=2.296$ (the members of the \CIV\ doublet are blended).
Hereafter, we refer to this
absorption system as system A. In the rest-frame of this system, the
SDSS spectrum covers the range 1153--2791~\AA\ while the UVES spectrum
covers the range 1091--2871~\AA. In the 2008 UVES spectrum, system A
has increased substantially in both strength and width.  Another
absorption system at $\sim5150\;$\AA\ ($z_{\rm abs}=2.329$; hereafter,
system B) also appears to have significantly increased in strength
between the two epochs. 
In the rest-frame of system
B, the SDSS spectrum covers the range 1142--2763~\AA\ and the UVES
spectrum covers the range 1080--2843~\AA. A complex system of much
narrower, associated \CIV\ absorption lines (hereafter, system C) is also present in the
range 5210--5270~{\AA} ($z_{\rm abs}=2.383$ for the whole complex, but several subsystems are identified, i.e., $z_{\rm abs}=2.374$, $z_{\rm abs}=2.385$ and $z_{\rm abs}=2.395$). These lines also appear to have varied between the
two observations. Other absorption systems (e.g., at 4960 {\AA}, 5050 {\AA}, and 5280 {\AA}) do not appear to have varied and they are unlikely to be related to the gas flow we are studying in this paper. 

\section{Analysis}
\label{sec:3}

\subsection{Continuum Normalization and Line Identification}
\label{sec:3.1}

\begin{figure}
\begin{center}
\includegraphics[width=9cm]{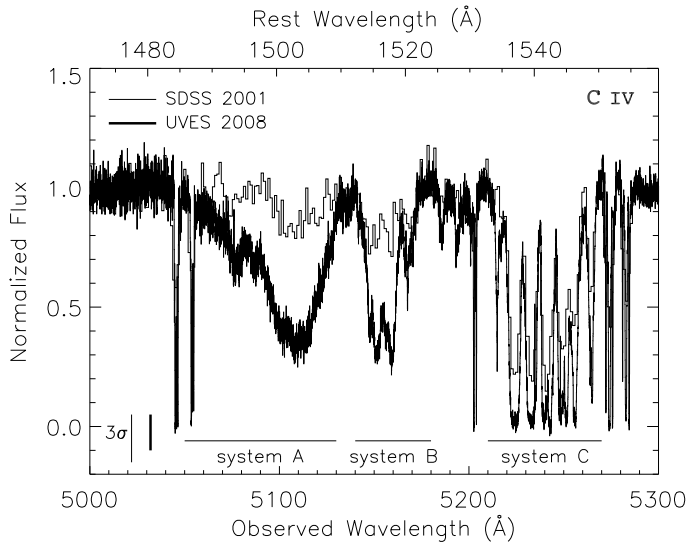}
{\vspace{0.5cm}} 
\caption[Figure2]{Normalized SDSS (thin) and UVES (thick) spectra
  around the \CIV\ absorption systems of interest. Horizontal lines
  indicate the studied \CIV\ absorption features: system A at
  $\sim$~5100 {\AA}, system B at $\sim$~5150 {\AA}, and system C at $\sim$~5240 {\AA}. Typical error
  sizes are represented on the bottom left corner. All three systems of study in the
  SDSS spectrum appear weaker than in the UVES spectrum. 
 }
\label{fig2}
\end{center}
\end{figure}


\begin{figure}
\begin{center}
\includegraphics[width=8.3cm]{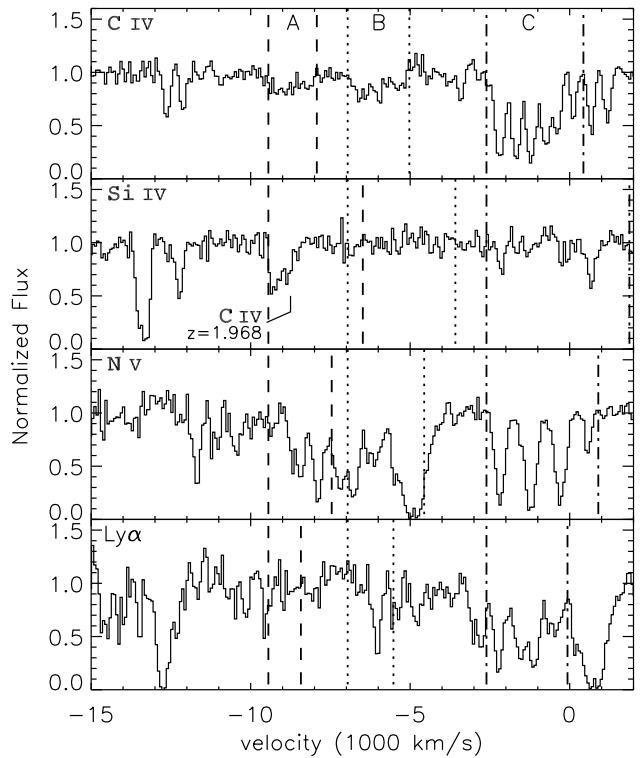}
\caption[Figure3]{Normalized SDSS spectrum on a velocity scale relative
  to the quasar emission redshift ($z_{\rm em}=$~2.399), showing the
  location of the \CIV\ doublet absorption and other possible transitions at the
  same velocity. The dashed lines represent the limits of system A
  in \CIV\ ($v_{\rm max,blue}= -9,520$ {\kms}), the dotted lines the
  limits of system B ($v_{\rm max,blue}= -6,960$ {\kms}) and the dashed--dotted lines the limits of system C ($v_{\rm max,blue}= -2,670$ {\kms}). These limits
  are based on the detected region for {\CIV} absorption, translated to cover the same regions (in velocity space) for both members of the {\SiIV} and {\NV} doublets, and for Ly{\a}. The absorption present in the {\SiIV} region for system A is an
  unrelated \CIV\ absorber at $z_{\rm abs}=$~1.968. The region where
  \NV\ might be present is too contaminated with lines in the Ly\a\ 
  forest to conclude whether it is present. In the case of {Ly\a}, the
  region is less contaminated, but it is not possible to assess
  whether the only absorption clearly present, in the region corresponding to system B at
  $v\sim$~5,900 {\kms}, is related Ly{\a} or due to an unrelated system.}
\label{fig3}
\end{center}
\end{figure}

\begin{figure}
\begin{center}
\includegraphics[width=8.3cm]{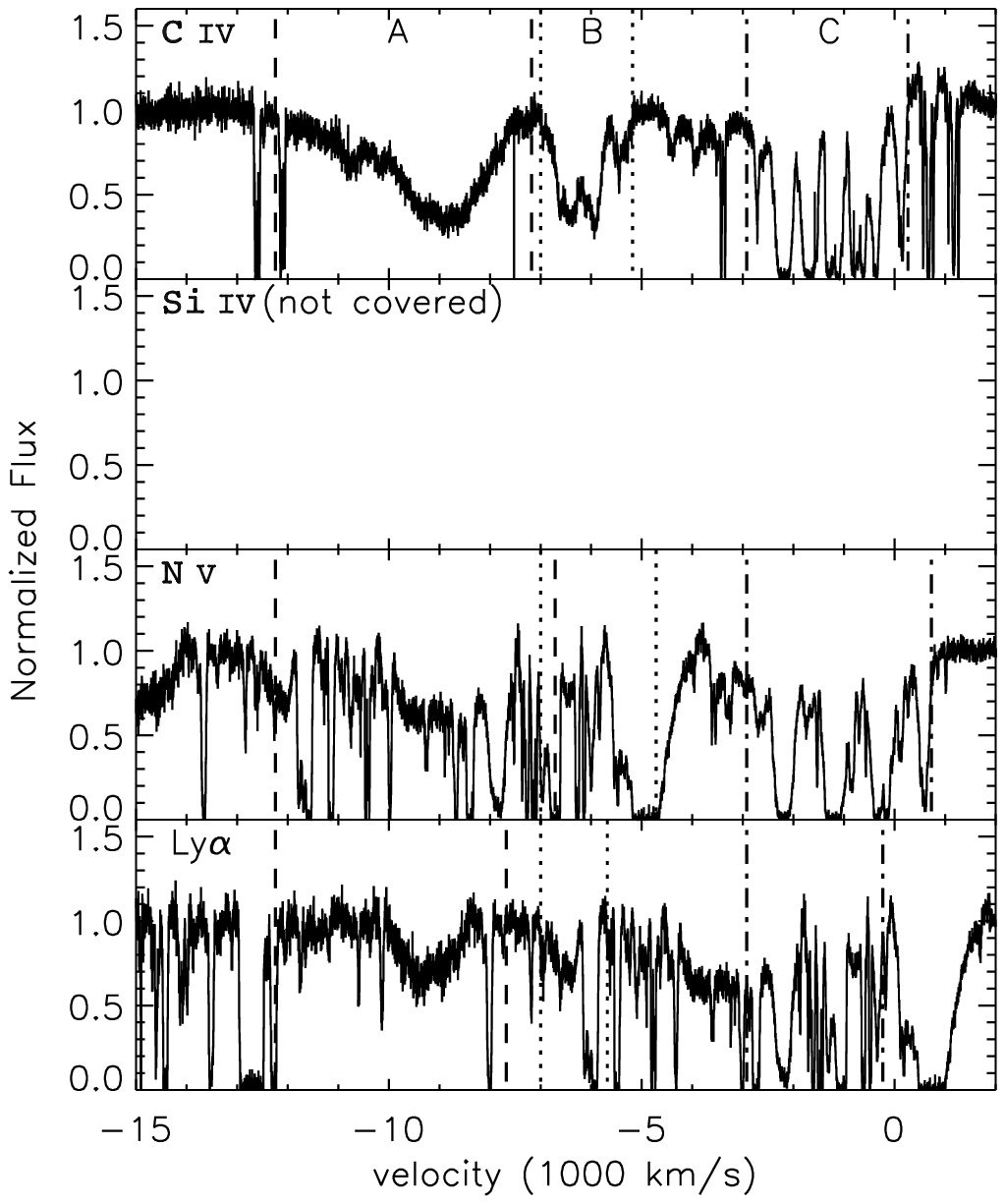}
\caption[Figure4]{Normalized UVES spectrum, similar to Fig.~\ref{fig3}. The dashed lines represent the limits of system A
  ($v_{\rm max,blue}= -12,240$ {\kms}), the dotted lines the limits of
  system B ($v_{\rm max,blue}= -7,000$ {\kms}), and the dashed--dotted lines the limits of the system C ($v_{\rm max,blue}= -2,870$ {\kms}). A gap in the echelle spectra leaves {\SiIV} uncovered. The region where {\NV} 
  might be present is, as in the the SDSS spectrum, quite contaminated
  by lines from the Ly{\a} forest
  (see \S\ref{sec:4.1}). {\NV} seems to be present in system A, probably in system B, and it is clearly present in system C. {Ly\a} appears in a less
  contaminated region, and absorption with the same kinematical
  structure as the {\CIV} absorption profile seems to be present in
  all systems A, B, and C.}
\label{fig4}
\end{center}
\end{figure}

To carry out measurements of the absorption troughs, we normalized the
SDSS and UVES spectra by the {\it effective} continuum, which
comprises of the true continuum and the broad emission lines.  To this
end we used cubic splines or low-order Chebyshev polynomials fitted to
portions of the spectrum that are free of absorption troughs. We
concentrated on producing a good fit to the effective continuum around
the absorption troughs of interest here, with the result that some of
the undulations elsewhere in the effective continuum are not perfectly
reproduced. The complex of associated absorption (system C), 
which removes a large portion of the {\CIV} emission line (see Fig.~\ref{fig1} around 5200--5300 {\AA}), as well as the Ly{\a} forest
lines, which appear blueward of 4200 {\AA},
make the normalization of certain wavelength regions difficult. 
We have taken a conservative approach, especially in the case of the UVES spectrum where the absorption lines are stronger, under-fitting the {\CIV} emission line where the choice is uncertain and using low-order polynomial when extrapolating its shape. The effective continua we used is shown in Figure~\ref{fig1} and the normalized spectra around the {\CIV} absorption lines
of interest are shown in Figure~\ref{fig2}.  
Our possible under-fitting of the {\CIV} emission line results in lower limits on the Rest-frame Equivalent Width ($W_{\rm rest}$) measurements for the complex of lines of system C (see \S\ref{sec:3.2}).
Similarly, we use a low-order polynomial when fitting the continuum regions of interest around other ions.
Note that differences in rest-frame equivalent widths
resulting from the choice of redshift measurement are considerably smaller than
the measurement uncertainties derived from the continuum placement. 

We searched the SDSS and UVES spectra for other absorption lines
commonly observed in the spectra of BAL and mini-BAL quasars at the
offset velocities of systems A, B, and C \citep[e.g., \OVIdbl, \HI\ Ly\a,
  \NVdbl, \OIlin, \SiIIlin, \CIIlin, \SiIVdbl, \SiIIdbl, \AlIIIdbl,
  \FeIIlinB, \FeIIlinR, \MgIIdbl; see][]{Hamann98, Arav01,
  RodriguezHidalgo11}. The absorption lines that we detected are shown
on a common velocity scale in Figures \ref{fig3} and \ref{fig4} (the
\SiIV\ region is missing from the UVES spectrum because of a gap in
spectral coverage, therefore the corresponding panel in
Fig.~\ref{fig4} is blank). 
The velocity scale in these Figures is set according
to the wavelength of the {\it blue} member of each
doublet. We adopt a convention in which
negative velocities denote a blueshift relative to the quasar rest
frame.
For reference, we mark the velocity windows of systems A,
B, and C in all the other panels as well, using, respectively, a pair of
dashed lines, a pair of dotted lines, and a pair of dashed-dotted lines. 
The maximum and minimum velocity of each region are defined based on a 3${\sigma}$ detection of the {\CIV} absorption. 
The blue end of each system is
always at the same velocity since it represents the maximum blueshift
of the {\it blue} member of a doublet dictated by {\CIV}, which is also the doublet
member used to set the velocity scale. The red end of each system varies
because it represents the maximum redshift of the {\it red} member of
a doublet and the separation of the doublet members is different for
different doublets.

None of the lower-ionization lines that are covered (those with rest-frame wavelength of approx.~1153--2791~\AA\ in the SDSS spectrum and 1091--2871~\AA\ in the UVES spectrum) were detected. In particular, {\MgIIdbl} is not covered in the SDSS spectrum and not detected in the UVES spectrum. {\OVIdbl} is not covered in either spectrum.
In the UVES spectrum shown in Figure~\ref{fig4} the combination of higher
S/N, resolution and stronger absorption lines allows us to better
discern the troughs in all systems in the {\NV} and Ly{\a} profiles.
Both transitions are clearly detected in several of the systems. 
In section~\ref{sec:4.1}, below,
we examine the absorption-line profiles seen in the UVES spectrum in
more detail, while in section~\ref{sec:3.2} we quantify the
variability via measurements of line strengths in both the SDSS and UVES spectra.

In the SDSS spectrum the {\SiIV} doublets of systems B and C are not detected.
The {\SiIV} doublet of system A may be present in the SDSS spectrum but
we cannot ascertain this because of contamination by an unrelated
{\CIV} system at $z_{\rm abs}=1.968$, which is marked in Figure~\ref{fig3}
\citep[part of a damped Lyman alpha system; see][] {Prochaska05}. Both
the \NV\ and Ly{\a} lines fall in the Ly{\a} forest of the quasar
spectrum. The numerous absorption lines in this region and the lower
S/N of this portion of the SDSS spectrum introduce additional
uncertainty in the placement of the effective continuum, as well as
the detection and measurement of absorption lines. Thus we cannot
confirm the presence of the {\NV} line of systems A or B in the SDSS
spectrum nor the Ly{\a} line of system A. The Ly{\a} line of system B
may be present but the observed absorption trough is more likely due to an intervening system (discussed in \S\ref{sec:3.2}). 
Both Ly{\a} and {\NV} show absorption in system C. 
Notice that the Ly{\a} of system C overlaps with the {\NV} from systems A and B, as we discuss in \S\ref{sec:3.2}.


\begin{figure}
\begin{center}
\includegraphics[width=8.3cm]{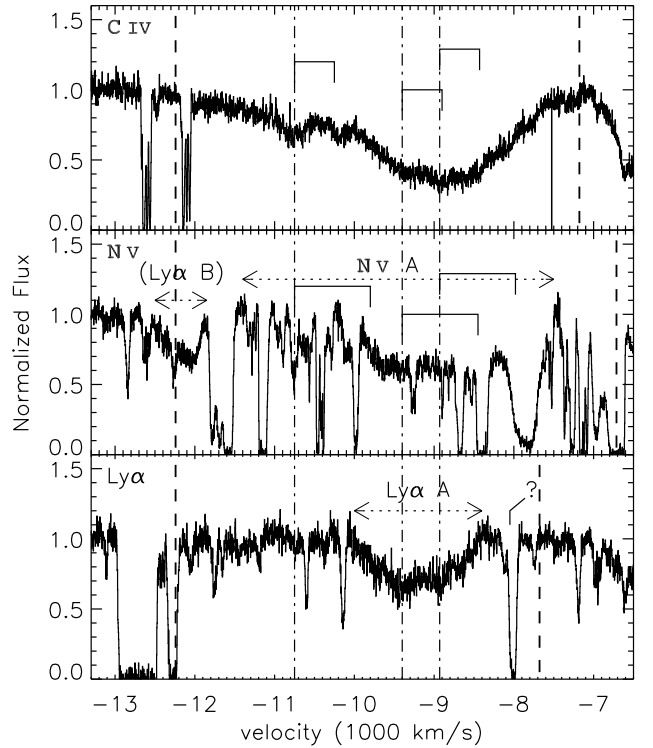}
\caption[Figure6]{SYSTEM A. Normalized UVES spectrum, similar to Figure \ref{fig4}, showing
  the location of the {\CIV} doublet absorption and other possible transitions at
  the same velocity for system A, which limits are represented by the
  dashed lines ($v_{\rm max,blue}= -12,240$ {\kms}). We include dashed--dotted
  lines to guide the eye about the location of the more prominent
  \CIV\ troughs and their correspondent \NV\ and Ly\a. Horizontal
  solid lines show the size of the doublet separations. The region where
  the different transitions are present is marked with dotted arrows. {\NV} is clearly present but blended with other lines and partially overlapping with the system B of {Ly\a}. {Ly\a}
  lies in a cleaner region and the resemblance to the {\CIV}
  absorption profile seems to confirm its detection. Unidentified lines are marked with ``?'' (see text).}
\label{fig6}
\end{center}
\end{figure}

\subsection{Profiles of Absorption Lines in the UVES Spectrum}
\label{sec:4.1}

\begin{figure}
\begin{center}
\includegraphics[width=8.3cm]{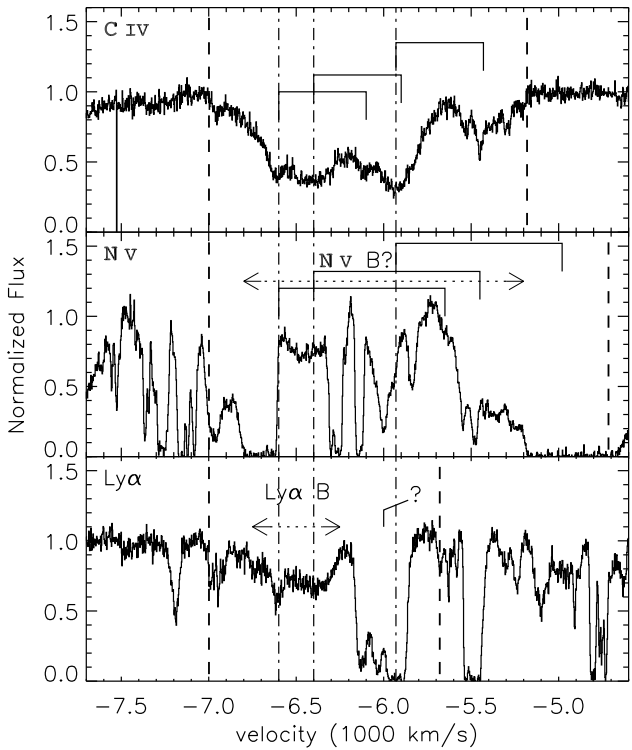}
\caption[Figure7]{SYSTEM B. Normalized UVES spectrum, similar to Figure \ref{fig6}, showing
  the location of the \CIV\ doublet absorption and other possible transitions at
  the same velocity for system B. Different line types and symbols are used as in Figure \ref{fig6}, with dashed lines representing in this case the
  limits of this system ($v_{\rm max,blue}= -7,000$ {\kms}). As in the case of
  System A (Fig.~\ref{fig6}), the spectrum is consistent with the
  presence of {\NV}, although it is severely blended with other unrelated lines. Ly\a\ shows a similar kinematic profile as the
  \CIV, where the location of the local minima appears at
  similar velocities between the two ionic species.
  }
\label{fig7}
\end{center}
\end{figure}

Figures~\ref{fig6} and \ref{fig7} show the UVES spectrum of systems A
and B, respectively. In each Figure we plot the {\CIV}, {\NV} and
Ly{\a} profiles on a common velocity scale with vertical dashed lines
showing the limits of the systems, as in previous Figures.

By inspection, we found minima in the complex and asymmetric {\CIV} profiles of
systems A and B and used them to identify possible kinematic
components.  We mark these with solid tick marks in Figures~\ref{fig6}
and \ref{fig7} (for the {\CIV} and {\NV} doublets we use pairs of tick
marks connected by bridges). In spite of the complexity of the line
profiles, the broad, asymmetric trough of system A is clearly seen in
Ly{\a} and is also discernible in {\NV}. The Ly{\a} profile of system A is broad ($\Delta v \sim$ 1700 {\kms})
resembling the kinematical structure of the corresponding {\CIV} system. There is also a narrower and deeper component ($\Delta v \sim 200\;${\kms}) at $v \sim -8,000\;${\kms} (marked as ``?''). It is unlikely to be related to the {\CIV} outflow of interest: it is black and does not have a counterpart in the {\CIV} profile.
The {\NV} profile of system A, though it shows some strong absorption, is
severely contaminated by the Ly{\a} forest and partly
overlaps with the Ly{\a} profile of system B ({\NV} at $v \sim
-11,600\;$\kms\ corresponds to the same spectral region as Ly{\a} at $v
\sim -5,900\;${\kms}) and system C ({\NV} at $v \sim
-7,800\;${\kms} corresponds to the same spectral region as Ly{\a} at $v
\sim -2,200\;$\kms; see Fig.~\ref{fig4}). As a result, we cannot be very confident that the
doublets and kinematic components identified in the {\CIV} and
Ly{\a} profiles have counterparts in {\NV}, although some absorption is clearly present. 

In the case of system B, the
{\CIV} doublets and kinematic components are easier to identify by
eye. 
As in system A, the {\NV} profile of system B is also severely
blended making it difficult for us to
verify the troughs that correspond to intrinsic kinematic components
seen in the other two transitions. 
The Ly{\a} profile (bottom panel), as in system A, resembles partially the {\CIV} profile in shape and 
velocity at the position of the strongest {\CIV} absorption. However, the kinematic
component of Ly{\a} at $v \sim -5,900\;${\kms} (marked as``?''), while it matches in velocity with a local minima in the {\CIV} absorption, is black and obviously
saturated, unlike any of the other components of Ly{\a} or other
transitions in this system. We discuss further the nature of the Ly{\a} and {\NV} absorption when exploring their variability in section~\ref{sec:3.2} and elaborate on the issue of saturation in section~\ref{sec:disc:properties}.

\subsection{Variability and Line Measurements}
\label{sec:3.2}

It is clear from a visual inspection of Figures \ref{fig2}, \ref{fig3}
and \ref{fig4} that the {\CIV} lines of systems A and B have become
considerably stronger between the 2001 and 2008 observations. 
The {\CIV} absorption profile becomes stronger over a velocity window
$\sim 12,520\;$\kms\ wide. The absorption troughs do not shift in
velocity and (1) the strongest parts of the troughs in the SDSS spectrum remain at approximately the same velocity in the UVES spectrum, and (2) regions of the SDSS spectrum that appeared unabsorbed are
part of the absorption trough seen in the UVES spectrum. The maximum
blueshift velocity of system A increases by $\sim
2,720\;${\kms} creating an asymmetric profile. This appearance of
additional absorption troughs in the same system without a change in
the velocity of existing troughs is characteristic of the variable
absorbers we have observed in our monitoring program and has also been
noted by other authors \citep[][Rodr\'iguez Hidalgo et al.\ in
  prep;]{Narayanan04, Misawa05, Hamann08, RodriguezHidalgo11,
  Vivek12}, The variability pattern described above suggests that the
region of the outflow responsible for the {\CIV} absorption is not
experiencing changes in velocity.

Comparing the {\NV} and Ly{\a} profiles between the two epochs is
complicated due to the multiple lines in the Ly{\a} forest. To
aid in this comparison, we overplot the SDSS and UVES spectra in
Figure 8, after smoothing the latter to the resolution of the
former, as we did in the top panel of Figure \ref{fig1}. We note that the strength of the
absorption troughs of both Ly\a\ and {\NV} in system A (between the
dashed lines, especially at $v\sim 9,500\;${\kms}) appear stronger/wider in
the UVES spectrum and so do the troughs of system B in
Ly{\a} (between the dotted lines, especially at $v\sim 6,500\;${\kms}) and system C in {\NV}. 
The {\NV} trough of system B, if present, does not appear to have
changed significantly.
We can also see that the sharp and black absorption trough in the
Ly{\a} profile in the UVES spectrum at $v\sim 8,000\;${\kms} (system
A; discussed in \S\ref{sec:4.1}) does not appear to be present in the older SDSS spectrum. On the other hand, the system at $v \sim -5,900$ {\kms} does not seem to have varied more than the 3{\s} error, which indicates that might be due to an unrelated Ly{\a} system. 
Similarly, the Ly{\a} trough of system C does not present significant variations between the two spectra. 
There is a partial overlap between the region corresponding to {\NV} 
and Ly{\a}, and the small variation at $v \sim -3,000$ {\kms} in Ly{\a} (bottom panel) is most likely to be attributed to {\NV} at $v \sim -8,500$ {\kms} (top panel).

\begin{figure}
\begin{center}
\includegraphics[width=8.8cm]{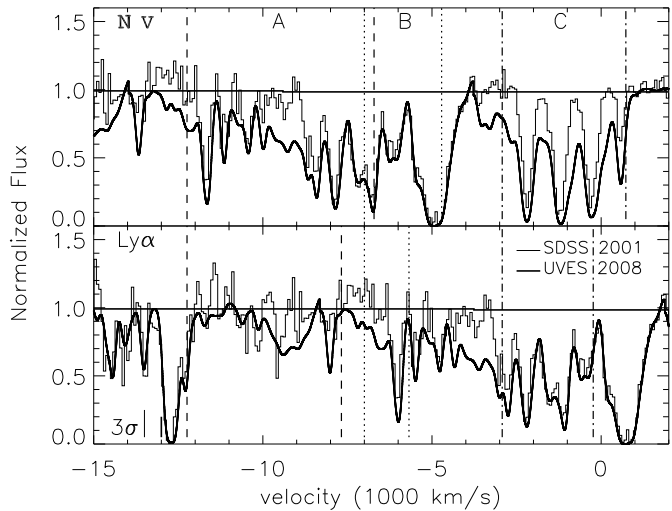}
{\vspace{0.25cm}} 
\caption[Figure5]{Normalized SDSS (thin line) and UVES (thick line)
  spectra in velocity scale relative to the quasar emission redshift
  ($z_{\rm em}=$~2.399), showing the location of the \NV\ doublet and
  Ly\a. To ease the comparison, we have smoothed the UVES spectrum to
  match the SDSS resolution. {\NV} has become stronger in systems A and C, but there is almost no variation between the two observations in system B.
  In the UVES spectrum, Ly{\a} is clearly present in system A at $v\sim$~9500 {\kms}, and a new component might have appeared at $v\sim$~8,000
  {\kms} (see text), while in system B we observe new absorption
  (at $v\sim$~6,500 {\kms}) and stronger previously observed narrow
  absorption (at $v\sim$~5,900 {\kms}), which might be due to an
  unrelated Ly\a\ system. System C seems to vary only in the region around $v \sim -3,000$ {\kms}, which is most likely due to {\NV}  at $v \sim -8,500$ {\kms} (see text).}
\label{fig5}
\end{center}
\end{figure}

In Table~\ref{table1} we include measurements of the rest-frame
equivalent widths ($W_{\rm rest}$) of {\CIV} in all systems and Ly{\a} lines in
systems A and B in the SDSS and UVES spectra. For the reasons we explain in \S\ref{sec:4.1} we do not attempt to
measure the {\NV} lines in either system or Ly{\a} in system C. We also give our best
estimates of the central wavelengths and maximum blueshift velocities
of the absorption troughs. We estimated the uncertainties in the
equivalent width measurements by repeating the continuum fits and
looking for the highest and lowest plausible placement of the
continuum level. These continuum adjustments are guided by the amount 
of noise around the absorption troughs of interest and do 
not modify the shape of the pseudo-continuum across the absorption 
profiles. The measurements in Table~1 reflect the increase in the
strength of the absorption lines described earlier. The equivalent
width of the {\CIV} line in system A has increased by a factor of
approximately 8, while that of the {\CIV} line in system B has increased
by a factor of approximately 2.5. While both measurements for system C
are included as lower limits due to the uncertain strength of the {\CIV} emission line, they also reflect the increase of strength
seen in the spectra.

We have also used the ``balnicity'' index and the absorption index to quantify the increase in the strength of the {\CIV} absorption \citep[BI and AI, respectively;
  see][]{Weymann91, Hall02, Trump06}.  We obtain ${\rm BI}=0\;$\kms, ${\rm
  AI}=850\;$\kms\ for the SDSS spectrum, and ${\rm BI}=590\;$\kms,
${\rm AI}=2,350\;$\kms\ for the UVES spectrum. The increase of each measurement is a result of the higher-velocity mini-BAL becoming broad enough to satisfy the formal definition of a BAL, with contiguous absorption deeper than 10\% below
the continuum level across a window of at least 1,000~\kms. 
We have also used a modified version of the AI that integrates the absorption trough from $v=-3,000\;$\kms\ to $-25,000$~{\kms} rather than from
$v=0\;$\kms\ as \citet{Trump06} did, to avoid the inclusion of the low velocity associated absorption of system C \citep[see
  discussion of issues associated with this convention
  in][]{Knigge08}. In this case, we obtain a value of ${\rm AI}=30\;${\kms} for the SDSS spectrum and ${\rm
  AI}=1,140\;$\kms\ for the UVES spectrum.
We note
that J1151+0204 was previously classified as a Hi-BALQSO by
\citet{Trump06} based on a measurement of ${\rm AI}=1,483\;$\kms (integrating from $v=0\;$\kms). This
measurement differs from ours because of a different continuum placement between our analysis and that of \citet{Trump06}.


\begin{table}
\scriptsize
\caption{Measurements of the Absorption features \label{table1}}
\begin{tabular}{lllrrl}
\hline
\hline
         &     &        & $v_{\rm max,blue}$ & $\lambda_{\rm central}$ & $W_{\rm rest}$ \\
spectrum & line & system & ({\kms})    & (\AA)             & (\AA) \\
\hline
SDSS & \CIV\ & A &  $-9,520$ & 5107.5 &  $1.1\pm 0.4$ \\
     &       & B &  $-6,960$ & 5153.5 &  $1.4\pm 0.3$ \\
     &       & C &  $-2,670$ & 5240.3 & $>$7.2 \\
\\
     & Ly\a\ & A &  $-9,520$ & \dots  &  $0.35\pm 0.09$ \\
     &       & B &  $-6,960$ & \dots  &  $0.25^{+0.09}_{-0.06}$ \\
\hline
UVES & \CIV\ & A & $-12,240$ & 5098.0 & $8.6^{+0.4}_{-0.7}$ \\
     &	     & B &  $-7,000$ & 5155.4 & $3.5^{+0.5}_{-0.1}$ \\
     &       & C &  $-2,870$ &  5240.1& $>$10.2 \\
\\
     & Ly\a\ & A & $-12,240$ & \dots  &  $1.41\pm 0.02$$\;^a$\\
     &       & B &  $-7,000$ & \dots  &  $1.82\pm 0.09$ \\
\hline
\end{tabular}    
\vbox{\medskip
$^a$ This value does not include the new narrow system
at $v \sim -8,000$ {\kms} (see \S\ref{sec:4.1}). This component has an
equivalent width of $W_{\rm rest} = 0.435\pm 0.003$ and, if included in the measurement, 
the relative increase in $W_{\rm rest}$ between
  the two epochs would be even larger. 
 }
\end{table}


\section{Discussion}
\label{sec:4}

\subsection{J1151+0204 in the Context of Previous BAL Variability Studies}
\label{sec:4.4}

In the preceding sections we described the transition from a
{\CIV} mini-BAL to a strong BAL in J1151+0204 on a rest-frame time
scale of $\lsim 2.86\;$years\footnote{All timescales in this section are converted to the rest-frame timescales of each quasar.}. The {\it appearance} of BALs in quasar spectra
that did not previously show them is a rare occurrence with only a
handful of such reports available in the literature, both in radio-quiet and radio-loud quasars.  {\it
High-velocity} \CIV\ BALs (with maximum velocities between $-25,000$
and $-50,000\;$\kms) have been reported to appear in a few quasars
that did not previously have them, namely J105400.40+034801.2, Ton~34,
and PG0935+417 \citep[see][]{Hamann08, Narayanan04,
Krongold10}. Upper limits on the time scale for these
transitions range from 1.5 to 8.3 years; in the case of PG0935+417 subsequent monitoring
showed that the variability continued on timescales of years \citep[][]{RodriguezHidalgo11}.  An
equally small number of cases of {\it low-velocity} \CIV, \SiIV, and/or
\MgII\ BALs (maximum velocities between $-6,000$ and $-12,000\;$\kms)
have also been reported, specifically in the quasars TEX~1726+344 and
J133356.02+001229.1 \citep[see][]{Ma02, Vivek12}. Upper limits on the
time scales for these transitions are 3.5 and 5.3 years
respectively. In an analogous event, BALs appeared in the far-UV
resonance lines of the Seyfert~1 galaxy WPVS~007, which previously
possessed only mini-BALs, on a rest-frame time scale of $\lsim
7\;$years \citep{Leighly09}. All of the above discoveries have been
fortuitous since there is no way to systematically select candidates
for this type of variability. Moreover, these discoveries often rely
on the comparison of new and archival spectra, which yields only large upper
limits on the variability time scale. Another consequence of the
sparse time sampling is that when the new BALs are discovered,
they are already fully formed, i.e., the intermediate steps of the
transition are not observed.

The case we have presented here is a transition from a mini-BAL to a BAL
in the same quasar. Because the absorption profiles appear at similar velocities and show similar kinematic structure in some of the systems, we interpret this transition as being most likely an
intermediate phase of the appearance of a BAL in a quasar without previous broad absorption lines. 
In fact, the time scale for the transition is very similar
to the time scales reported in the few previously known
cases. Similarly, the velocity of the BAL that has just emerged in
J1151+0204 is comparable to those of the low-velocity BALs cited
in the previous paragraph. Our present case is reminiscent of WPVS~007 \citep{Leighly09}
where the first set of observations revealed mini-BALs in the near-UV
resonance lines and the followup observations 7 years later showed the
appearance of BALs.  Moreover, it is possible that the emergence of
the \MgII\ BAL in J133356.02+001229.1, monitored by \citet{Vivek12},
represents a similar event to the one we have observed here, even
though an intermediate mini-BAL phase was not observed.  Taken
together, all the cases summarized above also suggest that mini-BALs
and BALs are connected to each other; possibly some mini-BALs are the
intermediate phases of the appearance or disappearance of BALs (see next paragraph), which 
would imply that quasars with mini-BALs and BALs might not be intrinsically different.
Dramatically varying mini-BALs, such as those reported by
\citet{Misawa05}, \citet{Hamann08, Hamann11}, and
\citet{RodriguezHidalgo11} could be related to this phenomenon,
corresponding to more prolonged and complex intermediate phases, or cases where the final profile never satisfies the definition of a BAL. Monitoring of similar cases at high spectral resolution will help discern whether the physical properties of the absorbers that produce mini-BALs are similar to those producing BALs.

In contrast to studies of the abrupt {\it appearance} of BALs, studies
of variability of existing BALs, including BAL {\it disappearance} can
be very systematic. A recent, extensive, systematic study of this type
has been carried out by \citet{FilizAk12} who have re-observed several
hundred BAL quasars discovered in the SDSS-I/II survey during SDSS-III
(probing rest-frame time intervals of approximately 1--4~years). They
find that a few percent of the \CIV\ BAL troughs in their sample
disappear in the interval between the two observations and they are
led to conclude that \CIV\ absorbers are detectable along our line of
sight for time intervals of order a century. The disappearing troughs
tend to be those with moderate equivalent widths and shallow
depths. They also observe that some of the troughs do not completely disappear but weaken into an intermediate mini-BAL phase, as in the case of J1151+0204. This is also seen in other quasars with multiple BAL troughs \citep[i.e.,][]{Hall11}, where some troughs are
observed to weaken while others disappear, reminiscent of the behavior
observed in J133356.02+001229.1 by \citet{Vivek12}.

\subsection{Inferring Some of the Properties of the Absorber in J1151+0204}
\label{sec:disc:properties}

The optical depths of the resolved {\CIV} doublets in system B appear
to depart from the 2:1 ratio expected for unsaturated
profiles (see the relative depths at $v \sim -6,500$ and
$-6,000\;$\kms\ in Fig.~\ref{fig7}). The depths of the lines approach
a ratio of 1:1, suggesting that the lines are saturated but the
absorber does not cover the background source completely \citep[a
  signature of ``partial coverage''; see][]{Barlow97, Hamann99,
  Ganguly99}. For fully resolved, unblended profiles, one can use the
relative depths of the doublet to derive the coverage fraction as a
function of velocity, $C_f(v)$, defined as the fraction of photons
from the background source that pass through the absorber. This
derivation relies on the assumption that the absorber is homogeneous,
(i.e., the optical depth as a function of position across the face of
the background source is uniform)\footnote{This need not be the case
  in practice as measurements of different coverage fractions in
  different lines suggest that the absorber may not be homogeneous;
  \citep[see, for example][]{Arav08}. Nonetheless, we are not able to
  constrain the relevant properties of the absorber here, therefore we
  make this simplifying assumption and examine the consequences
  \citep[see also the discussion in][]{Hamann11}.}.  In the special
case we are considering here, where the lines are fully saturated but not
black, the coverage fraction, as a function of velocity across the
profile is given by $C_f(v) = 1 - I(v)$, where $I(v)$ is the
normalized intensity in the observed absorption troughs.  Thus, we
arrive at $C_f(v)\approx 0.65$ as a plausible estimate in the velocity
range from $-6,700$ to $-6,300\;$\kms\ in system B. This is subject to
the caveat that the {\CIV} troughs probably comprise multiple blended
kinematic components, as indicated in Figure~\ref{fig7}. Assuming that
the {\CIV} line in system A is also saturated we then infer the same
value of $C_f(v)$ as for system B, based on the fact that its depth is
approximately the same as that of system B (see Figs.~\ref{fig2} and
\ref{fig6}).

Since the {\CIV} absorption troughs of systems A and B  are superposed on the continuum and
not on the {\CIV} emission lines, we can adopt a simple picture in
which the absorber covers only the UV continuum source. Thus we can
constrain the transverse size of the absorber by comparing it with the
size of the continuum source, under the assumption that the absorber
has a uniform optical depth and sharp edges (i.e., partial coverage is
a simple geometrical effect). To this end we first estimate the mass
of the central black hole following the three different prescriptions
of \citet{Warner03}, \citet{Vestergaard06}, and \citet{Shen12}.  We
make use of the FWHM of the {\CIV} emission line (none of the other preferred estimators, such as {\MgII} and H$\beta$, are covered in our spectra) and the continuum
  luminosities at rest wavelengths of 1350~\AA\ and 1450~\AA, assuming
  a cosmological model with $H_o = 71$ \kms\ Mpc$^{-1}$, $\Omega_M = 0.27$,
  $\Omega_{\Lambda}=0.73$.  The method employing the FWHM of the
  {\CIV} emission lines is subject to the caveat that the blue wing of
  the line is contaminated by the associated absorption complex
  discussed in earlier sections. We arrive at $M_{\bullet}
    \approx 3 \times 10^9 M_{\sun}$. Although the \CIII] emission line is also present, it is contaminated by the {\SiIII}{\lb}1892 and {\AlIII}{\lb}1857 emission lines and the black hole masses derived
    from this method can deviate significantly from those derived
    using the FWHM of the H$\beta$ line. To determine the effective size
    of the continuum source at a rest wavelength of 1510~\AA\ we
    follow the method described in \citet{RodriguezHidalgo11}, based
    on \citet{Peterson97}. This method relies on the \citet{Shakura73}
    model of the disk (assuming a radiative efficiency of $\eta =0.1$)
    and makes use of a bolometric luminosity of $L_{\rm bol} =
    1.1\times 10^{47}\;{\rm erg\; s}^{-1}$ \citep[obtained from
      $L_{\rm bol}\approx 4.36\,\lambda L_{\lambda}(1450 {\rm \AA})$;
      see][]{Warner04}. The final result was boosted by a factor of 4
    to account for the fact that quasar microlensing studies
    systematically yield a larger continuum source size than what the
    \citet{Shakura73} model predicts \citep[see][]{Morgan10}.  We
    arrive at a continuum source diameter of $d_{\rm cont}\approx
    8\times 10^{16}\;$cm. To assess the sensitivity of this result to
    the uncertainty in the black hole mass, we varied the black hole
    mass by an order of magnitude in either direction and found a
    corresponding change in the diameter of the continuum source by a
    factor of 2. Assuming that the projected area of the absorber is
    smaller than that of the continuum source and that the coverage
    fraction infered earlier represents the fraction of the area of
    the continuum source covered by the absorber, we obtain a
    characteristic transverse dimension of the absorbing gas at $-6,700$ to $-6,300\;${\kms} of, at least, $\sim
    6\times 10^{16}\;$cm.
    
In contrast, the narrower absorption line complex (system C) at lower velocities is saturated and black, at least in the UVES spectrum\footnote{There seems not to be significant change in the {\CIV} emission line profile between the two epochs, as shown in Figure \ref{fig1} where both spectra are over-plotted after a ``normalization'' that did not include the emission lines.}, suggesting that, if no other broad, shallow absorption component is present, the absorber covers completely the broad emission line region and the continuum.  This system also increases in strength between the two observations, suggesting that these lines are intrinsic to the quasar environment. They also show larger widths compared to other associated lines (for example, those at $\sim$~5050 {\AA}; see Fig.~\ref{fig2}), which is characteristic of intrinsic absorption \citep[][]{Vestergaard03}. Moreover, systems A, B, and C are all varying in concert, which suggests that they are part of the same outflow and subject to the same influences (e.g., the same ionizing continuum). While we can affirm that this absorber is more extended in size than the system discussed in the previous paragraph (due to the larger coverage fraction), the spatial relation between this complex system and the outflowing gas of system A and B is unknown. 

The gas responsible for the Ly\a\ absorption in systems A and B may also
cover the background source partially since the Ly\a\ profile is
similar to the {\CIV} profile (see Fig.~\ref{fig6}). In this case,
however, we cannot make a robust estimate of the coverage fraction but
instead we can obtain upper and lower bounds of $0.42 < C_f(v) < 1$
based on the depth of the trough in system A. If the black, saturated
trough at $v = -5,900\;$\kms\ in system B is part of the
Ly\a\ profile, then the $C_f(v)$ likely spans a range of values across
the Ly\a\ profile.
We would obtain a lower limit on the hydrogen column density of
$N({\rm Ly}\alpha) > 4.8 \times 10^{14} \;{\rm cm}^{-2}$, by assuming
that the Ly\a\ trough in system A is not saturated and it fully covers
the background source (i.e., $C_f=1$). However, Ly\a\ saturates at, approximately, $10^{14} \;{\rm cm}^{-2}$ and it would be very difficult to get unsaturated Ly\a\ with the amount of observed {\CIV}. Thus, using the value of $C_f=0.42$ we obtain a better estimate of $N({\rm Ly}\alpha) \gsim 1.6 \times 10^{15} \;{\rm cm}^{-2}$. No lines from lower ionization species (such as \MgII, \FeII, and \AlIII) were detected in either
epoch. It is not surprising to detect saturated Ly{\a} in the same phase as {\CIV} and {\NV} \citep[see, for example,][]{Wu10b}, and the absence of other low ionization species supports the idea that these three lines originate in the same ionization phase.

\subsection{Possible Causes of the Observed Variability}
\label{sec:4.3}

The observed variability can be a result of, at least, one of the following effects: 
variations caused by changes in the ionization state
of the absorber and variations caused by motion of the absorber across
the cylinder of sight; a combination of the two is also possible. The changes in the line profiles expected from
each of these two extreme scenarios are illustrated in \citet[][]{Misawa05}.
The fact that we do not observe any substantial change in the apparent
outflow speeds of the absorption troughs is consistent with both of
these interpretations. Therefore, we examine each of these two
possibilities in turn.

The ionization state of the absorbing gas can change as a result of
fluctuations in the ionizing continuum that illuminates it. In this
context we can obtain a lower limit on the electron density in the
absorbing gas by requiring that the recombination time of the
\CIV\ ion is shorter than the interval between observations
\citep[see][]{Hamann97a}. This yields $n_e\gsim (\alpha_r\, \Delta
t_{\rm rest})^{-1} \approx 4,000\;{\rm cm}^{-3}$ \citep[where $\alpha_r$
  is the recombination coefficient with values of $5.3\times 10^{-12}$
  and $2.8\times 10^{-12}\;{\rm cm^{-3}\; s^{-1}}$ for
  \CIII$\rightarrow$\CIV\ and \CV$\rightarrow$\CIV, respectively;
  taken from][assuming a temperature of 20,000~K; we use the value for {\CV}$\rightarrow${\CIV} since it results on the strongest limit]{Arnaud85}. 
Moreover, because we observe {\NV} and Ly{\a} increasing as well as {\CIV}, in a scenario where these changes are due to ionization variations it is more likely that the ionization parameter is decreasing \citep[see Appendix A in][]{Hamann11}.  
This scenario relies on the dominant ionization state of C changing from
 \CV\ to \CIV\ on a time interval shorter than 2.86~years,
hence the intensity of the ionizing continuum must change by at least
an order of magnitude on this time scale\footnote{A more sophisticated approach to infering the density from the
variability time scale is given in \citet[see their
equation 10]{Arav12}. However, that approach can only be applied when more
detailed information than what we have here is available about the
variations of the ionizing continuum.}. We consider this an unlikely
possibility for such a luminous quasar, which should vary quite
slowly. In the sample of luminous quasars monitored by
\citet{Kaspi07}, for example, the most extreme variation observed is
the brightening of SBS~1116+5603 by 40\% in a rest-frame time interval
of 3~years. Even the moderate-luminosity PG quasars monitored by
\citet{Giveon99} do not vary by more than 60\% over rest-frame time
intervals of about 5 years. However, it is possible that the flux
illuminating the absorber can change by large factors as a result of
variable transmission through a porous or patchy ``screen'' between the
absorber and the continuum source. This possibility was suggested by
\citet{Misawa07b} and it is supported by observations of rapid,
high-amplitude X-ray variability in mini-BAL quasars
\citep[e.g.,][]{Giustini11}. Such a scenario is also compatible with the observation that multiple absorption systems (A, B, and C) at different velocities and with different trough shapes vary in concert. 

If, instead, we suppose that the variations in the absorption troughs
are the result of a parcel or filament of gas entering the cylinder of
sight, then the transverse speed of the parcel can be estimated by
assuming that the parcel has traversed the diameter of the continuum
source in the interval between our two observations, i.e. $v_{\rm
  trans} \gsim d_{\rm cont} / \Delta t_{\rm rest}\approx
8,900\;$\kms. If we further assume that the parcel is moving at the
local dynamical speed (either because it is in Keplerian rotation or
because it has been launched at the local escape speed; this
assumption is further justified by the similarity of the transverse
speed to the apparent outflow speed), then we obtain the following
constraint on its distance from the black hole: $a \lsim GM_{\rm
  \bullet}/v^2_{\rm trans} \approx 5\times 10^{17}\;{\rm cm} \approx
1100\;r_{\rm g}$ (where $r_{\rm g}=GM_{\bullet}/c^2$ is the
gravitational radius). This distance places the absorber in the
vicinity of the broad-emission line region \citep[hereafter BELR;
  cf,][]{Kaspi05, Kaspi07} and suggests a connection between the
emission- and absorption-line gas.

Future monitoring at high resolution of the multiple ions present in the outflow is necessary to better characterize the variability and, thus, find its cause. The simulations of these processes presented in \citet{Wise04} and \citet{Misawa07b} can form the basis for the observational tests.

\section{Summary and Conclusions}
\label{sec:5}

We present the discovery of dramatic variability in the
\CIVdbl\ mini-BAL of the radio-quiet quasar J115122+020426 ($z_{\rm
  em}=$~2.399).  Our results are based on the comparison of an older
SDSS spectrum and a newer VLT/UVES spectrum.  The original mini-BAL
profile had a FWHM of 1,220~\kms\ and a maximum blueshift velocity of
9,520~\kms. Over a rest-frame time interval of 2.86 years, the
equivalent width increased by a factor of 8 and the maximum blueshift
velocity became 12,240~\kms. There is no clear evidence for
changes in the centroid velocity of the absorption troughs. Instead the
absorption lines in the earlier spectrum became stronger and regions
of the spectrum that appeared unabsorbed in the earlier spectrum are
clearly absorbed in the later spectrum. The properties of the new
absorption trough satisfy the definition of a BAL. Thus we have
caught a rare transition of a mini-BAL to a BAL. Two other absorption
systems at somewhat smaller blueshift velocities (one of them a
cluster of narrow absorption troughs) also became substantially
stronger in the same time interval.  We observe a similar variability
pattern in the \NV\ and Ly\a\ absorption lines.

From the residual depths of the lines in the \CIV\ mini-BAL (system B) we infer
that the absorber covers a fraction $C_f = 0.65$ of the area of the
background continuum source. In contrast, the absorber responsible for
the cluster of narrow \CIV\ lines appears to cover the background
continuum source completely during the later observation. If we
ascribe the variability to changes in the coverage fraction due to
motion of the absorber at the dynamical speed, we conclude that the
absorber is located at a distance from the continuum source comparable
to the size of the BELR. If, on the other hand, the variability is a
result of changes in the ionization state of the absorber, the short
variability time scale suggests that the ionizing flux incident on the
absorber is modulated by a screen with rapidly variable transmission.

Future monitoring is necessary to better characterize the cause of
this variability, in particular to determine whether the variability
is caused by motion of the absorber across the cylinder of sight or
changes in its ionization state. The simulations of these processes
presented in \citet{Wise04} and \citet{Misawa07b} can form the basis
for the observational tests.

\section*{Acknowledgments}

We thank the anonymous referee for thoughtful comments.
This research was funded by NASA under grant NAG5-6399 NNG04GE73G, and
by the National Science Foundation (NSF) under grants AST-0407138 and
AST-0807993 and through the REU program. P.R.H. would like to thank
Pat Hall for fruitful discussions and comments on the manuscript. P.R.H. is currently supported by an Ontario
Early Researcher Award to P.B.H. M.T.M. thanks the Australian Research Council for a QEII Research Fellowship (DP0877998).

Funding for the SDSS and SDSS-II has been provided by the Alfred
P. Sloan Foundation, the Participating Institutions, the National
Science Foundation, the U.S. Department of Energy, the National
Aeronautics and Space Administration, the Japanese Monbukagakusho, the
Max Planck Society, and the Higher Education Funding Council for
England. The SDSS Web Site is http://www.sdss.org/.

The SDSS is managed by the Astrophysical Research Consortium for the
Participating Institutions. The Participating Institutions are the
American Museum of Natural History, Astrophysical Institute Potsdam,
University of Basel, University of Cambridge, Case Western Reserve
University, University of Chicago, Drexel University, Fermilab, the
Institute for Advanced Study, the Japan Participation Group, Johns
Hopkins University, the Joint Institute for Nuclear Astrophysics, the
Kavli Institute for Particle Astrophysics and Cosmology, the Korean
Scientist Group, the Chinese Academy of Sciences (LAMOST), Los Alamos
National Laboratory, the Max-Planck-Institute for Astronomy (MPIA),
the Max-Planck-Institute for Astrophysics (MPA), New Mexico State
University, Ohio State University, University of Pittsburgh,
University of Portsmouth, Princeton University, the United States
Naval Observatory, and the University of Washington.

\bibliographystyle{apj_modified}
\bibliography{bibliography.bib}

\end{document}